# An energy efficient service composition mechanism using a hybrid meta-heuristic algorithm in a mobile cloud environment

Godar J. Ibrahim[1]*, Tarik A. Rashid[2], and Mobayode O. Akinsolu[3]

[1] Department of Software and Informatics Engineering, Salahaddin University-Erbil, Erbil, Iraq

*Corresponding author's Email: godar.ibrahim@su.edu.krd

[2] Computer Science and Engineering Department, University of Kurdistan Hewlêr, Iraq

Email: tarik.ahmed@ukh.edu.krd

[3] Faculty of Arts, Science and Technology, Wrexham Glyndwr University, Mold Road, Wrexham, LL11 2AW UK

Emial: mobayode.akinsolu@glyndwr.ac.uk

**Abstract**

By increasing mobile devices in technology and human life, using a runtime and mobile services has gotten more complex along with the composition of a large number of atomic services. Different services are provided by mobile cloud components to represent the non-functional properties as Quality of Service (QoS), which is applied by a set of standards. On the other hand, the growth of the energy-source heterogeneity in mobile clouds is an emerging challenge according to the energy-saving problem in mobile nodes. To mobile cloud service composition as an NP-Hard problem, an efficient selection method should be taken by problem using optimal energy-aware methods that can extend the deployment and interoperability of mobile cloud components. Also, an energy-aware service composition mechanism is required to preserve high energy saving scenarios for mobile cloud components. In this paper, an energy-aware mechanism is applied to optimize mobile cloud service composition using a hybrid Shuffled Frog Leaping Algorithm and Genetic Algorithm (SFGA). Experimental results capture that the proposed mechanism improves the feasibility of the service composition with minimum energy consumption, response time, and cost for mobile cloud components against some current algorithms.

# 1. Introduction

Lately, mobile cloud computing has been achieving significant progression. The rapid development of cloud computing, mobile devices, and Internet of Things (IoT) domains [1, 2] have managed to considerable growth in the number of near-feasible web services that have different Quality of Service (QoS) metrics. The service composition method as an NP-hard problem consists of creating a new collection of atomic services for providing user requirements as Service Level Agreement (SLA) assurance [3].

The QoS metrics are classified into two groups: provider dependent and user-dependent features [4, 5]. The QoS metrics of user-dependent like price and popularity have identical values and are evaluated at the user side [6, 7]. On the other hand, provider dependent QoS metrics like energy consumption and throughput are assessed at the provider side. According to apply acceptable services from mobile cloud providers to user's requirements, energy efficiency is a critical issue in mobile cloud service composition [8].

For energy-saving effort, refining energy-aware effectiveness has developed as an efficient despicable to cover mobile cloud providers, such as cellphones, and IoT devices using the power-harvesting technology [9]. Due to its complexity of realizing interoperability between different mobile nodes [10], the service composition methods are selected by global optimization in which aggregated-quality are maximized and global constraints are preserved. Due to the restriction of energy-saving resources in the mobile nodes, such as battery life and power-saving level, the mobile cloud providers should be recognized to present an optimal selection of atomic services, in which each mobile node can save battery life for other services. In spite of the restriction of power-saving and energy consumption problem in the mobile cloud providers, finding an optimized service composition mechanism with consuming low energy factors is a key challenge.

In this paper, we proposed a hybrid Shuffled Frog Leaping Algorithm (SFLA)[11] and Genetic Algorithm (GA)[12] (SFGA) to minimize the energy consumption of mobile cloud providers for composing required atomic services for achieving optimal QoS criteria. The main contributions of this research are as follows:

1) Proposing an energy-aware service composition mechanism to minimize the energy consumption of mobile cloud providers.

2) Presenting a hybrid SFGA algorithm for the proposed mobile cloud service composition mechanism.

3) Providing optimal solutions for the proposed service composition mechanism based on the maximum level of QoS factors.

4) Evaluating the proposed service composition mechanism using the SFGA algorithm in three scenarios.

The rest of this research is organized as follows: In Section 2, some relevant research studies are discussed. Section 3 presents the problem statement on the mobile cloud service composition. Section 4 shows the proposed method with details of the hybrid meta-heuristic algorithm. Section 5 represents the simulation results based on three scenarios to analyze and assess the feasibility of the proposed method. Finally, Section 5 illustrates the conclusion and future.

## 2. Related work

In this section, some of the substantial and recent approaches are introduced. The QoS-aware service selection method in [13] is presented, which uses a new combination of Tabu search and hybrid Genetic algorithms to optimize the service selection problem. Experimental results of the study were compared with the basic genetic algorithm (BGA) and iterative steepest descent method (ISD) approaches. They claimed which the gained results are shown that the proposed hybrid genetic algorithm betters the performance of the mentioned techniques. But, it is less stable behavior than HGA in large scale problem and it is not a proper approach in large scale problems.

In [14], integral meta-processes and a conflict detecting mechanism (CDM) are designed to facilitate for process designers. Conflicts are defined as events that cause accurate meta-process design and could violate the normal or expected results. The process design gap and the systematic designing guideline as two important issues are fixed by the proposed algorithm. In the process design gap issue, to use web-based services, the gap between traditional process design and process integration for enterprise migration is investigated. On the other hand for systematic designing guidelines, a systematic step-by-step Conflict Detecting Mechanism (CDM) is proposed to solve this problem. In the mentioned method, conflict detecting operations are divided into a three-phased process analysis: The validation, verification, and performance analysis phases. The validation and verification phases try to build confidence in which system is appropriate to achieve the considered aims. Also, the performance analysis phase is exploited to investigate feasible conflicts and also examine for process execution times, which could violate the aims.

A few researchers proposed a quality constraints decomposition (QCD) method, which uses a top-down method to decompose the global constraints into the local constraints [15]. To do this, the presented approach applies to the genetic algorithm (GA). In addition, a simple linear search is used to search the best web service for each task in the proposed algorithm. Contrary to current approaches, the QCD approach is not complex due to the use of a limited set of tasks, particularly in the case of dynamically distributed service composition. The proposed algorithm experiments on the QWS dataset and the obtained results illustrated that the QCD method improves the quality of services QoS i.e. computation time concerning the number of web services.

In the work [16], for supporting the dynamic adaptation of service compositions, an approach based on a semantically rich variability model was introduced. The pointed method is forced to make a decision when a problem occurs in the context. The composition model is affected and changed by the activation and deactivation of features in the variability model, which breaks down the principle of service composition. The service composition process, especially at the runtime, is changed by adding or removing fragments of the Business Process Execution Language (WS-BPEL) code. Thus, they used Constraint Programming to verify the variability model and its possible configurations at design time. Experimental results demonstrated that the proposed method has several advantages at design time and runtime.

A nonlinear optimization model with constraints for the services selection problem was proposed in the first step [17]. Afterward, a new discrete invasive weed optimization algorithm was proposed for the web services selection process. The suggested method runs in two phases. In the first phase, a collection of practical solutions is produced randomly and then they are encoded into decimal code. In the next phase, to spread solutions in the problem space Gaussian diffusion is employed to guide the population. Furthermore, when the standard deviation of Gaussian distribution is changed then the mutation probability and mutation step size of an individual is adjusted dynamically. Hence, the diversity of the population is guaranteed in the first phase and the approach is enabled to cover the search space, while the local search nearby excellent. On the other hand, global convergence of the individuals is considered in the second phase. Finally, the experiment results and analysis had proven the robustness and feasibility of the proposed algorithm.

In the study [18], for computing end-to-end QoS values of vertically composed services, a prediction model was proposed, which is based on the Software as a Service (SaaS), Infrastructure as a Service (IaaS) and Data as a Service (DaaS). The QoS values are used to predict the service selection and recommendation process. Historical QoS values and cloud service and user information are used by the proposed model for predicting unknown end-to-end QoS values of composite services. The experimental results revealed that the proposed model bettered prediction accuracy. Also, the impact of different parameters on the prediction results was studied. The real cloud services QoS dataset was used in the experiments.

In order to select fittest candidates' services with regard to QoS aspects, many-objective evolutionary algorithms are explored to address the selection of web services based on a real-world benchmark by considering nine QoS properties [19]. Experimental outcomes were illustrated that the newly presented algorithm had a proper trade-off among the QoS properties. Also, the results of the experiment showed that the computational cost of the proposed algorithm is reasonable.

The authors in [20], presented a systematic review of formal modeling and analysis of service composition approaches based on the evaluation of functional and nonfunctional specifications. A technical comparison and discussion were applied to evaluate existing challenges on the service composition approaches.

A multi-attributee transition model to prove the correctness of a service composition method in a multi-cloud computing environment based on event-based QoS factors was presented in [21]. Some functional properties were analyzed to illustrate the correctness of the proposed method.

Finally, Sun, et al. [22] presented an IoT service composition approach based on the timed-concentrated mechanism in a large scale task environment. To evaluate the proposed approach, the authors compared some metaheuristic algorithms to check and analyze the performance of the proposed service composition approach in the IoT environment.

## 3. Proposed hybrid algorithm

This section presents a service composition mechanism based on a hybrid meta-heuristic algorithm. First, the service composition model is presented in subsection 3.1. Second, we illustrate the proposed algorithm by using SFGA for mapping into the mobile service composition mechanism in subsection 3.2. Finally, the fitness function for the proposed algorithm is described in subsection 3.3.

## 3.1 Service composition model

This section presents the service composition approach based on a hybrid algorithm. The complex tasks of users based on the algorithm [23, 24] are converted into *m* abstract tasks ($Task_1$, $Task_2$, ...$Task_m$) [25]. For each $Task_i$, there is an abstract mobile service provider $S_i$ that has the same functionality and different QoS values. $S_i$ is a service provider that contains existing mobile services including $ws_{i,j}$ to represent the user.

The QoS values are shown with $q(ws_{i,j}) = (q_1, q_2, ..., q_r)$, where $q_r$ illustrates the appraisal value of the $r^{th}$ feature of mobile web service $ws_{i,j}$. Individual services need to be collected into a composite service *cs* to meet the increasingly complex demand and provide an aggregation function [26].

Figure 1 presents the service composition process to select optimal candidate services with the highest QoS values in the following steps [27]. In the first step, the service discovery is applied according to different QoS factors for *m* abstract tasks. In step 2, service selection between $WS_{1,1}$ ... $WS_{m,n}$ is specified to find a composition path based on QoS factors. Step 3 illustrates aggregating and evaluating the QoS values of each composite path according to the service composition workflow. Finally, Step 4 shows selecting maximum QoS values for the optimal composition solution between all the candidate composition paths [28].

## 3.2. Proposed hybrid service composition model

This study represents a hybrid service composition approach by using SFLA and GA optimization algorithms. The ordinary SFLA algorithm is designed to optimize the continuous problems. Therefore, it is not possible to apply for a service composition issue as a discrete problem[29]. To do this, the encoding of individuals is established to have proper and valid solutions in the applied SFLA. Also, the GA algorithms are embedded in SFLA to evolve the individuals. In the following, firstly, the use case scenario for the service composition will be presented and then the applied evolutionary algorithms will be described in detail.

### 3.2.1. Use case scenario and Algorithms for the proposed service composition

The main functions of the service composition system are illustrated by the use case diagram in Figure 2. In this scenario, the use case diagram has two actors, namely user and service provider. According to the diagram, the system contains two subsystems: service composition and process results. 'QoS needs', 'Service request', 'Interpret information', 'Analyze service', 'Service discovery', 'Service Composer', and 'Service delivery' are the use cases, which are shown in the use case diagram. First, the user requests services according to his QoS needs. Then, the entered information is interpreted. Afterward, the required services are discovered and analyzed. Finally, the services with high QoSs are composed and delivered to the user.

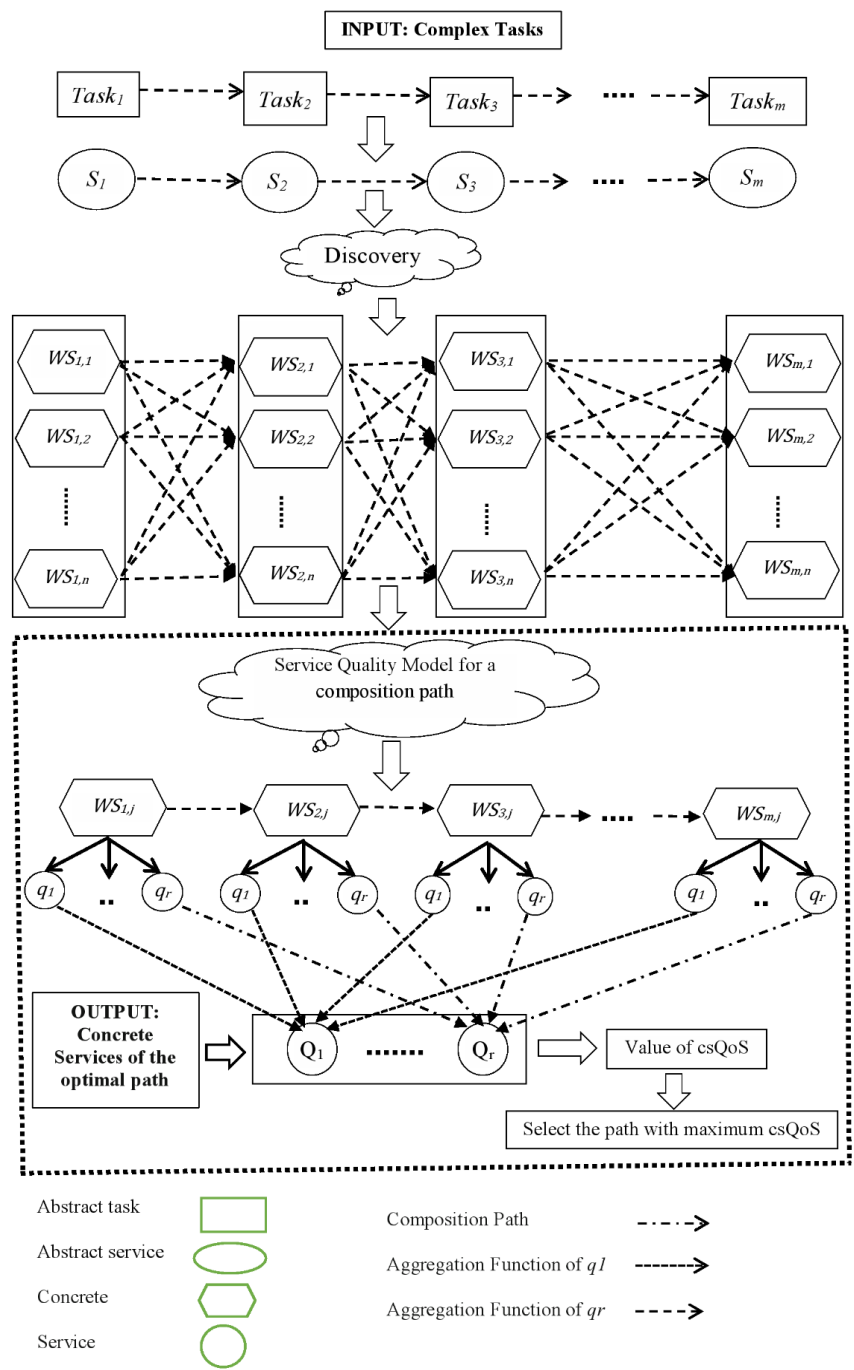

**Figure 1. The proposed service composition process in cloud computing [30].**

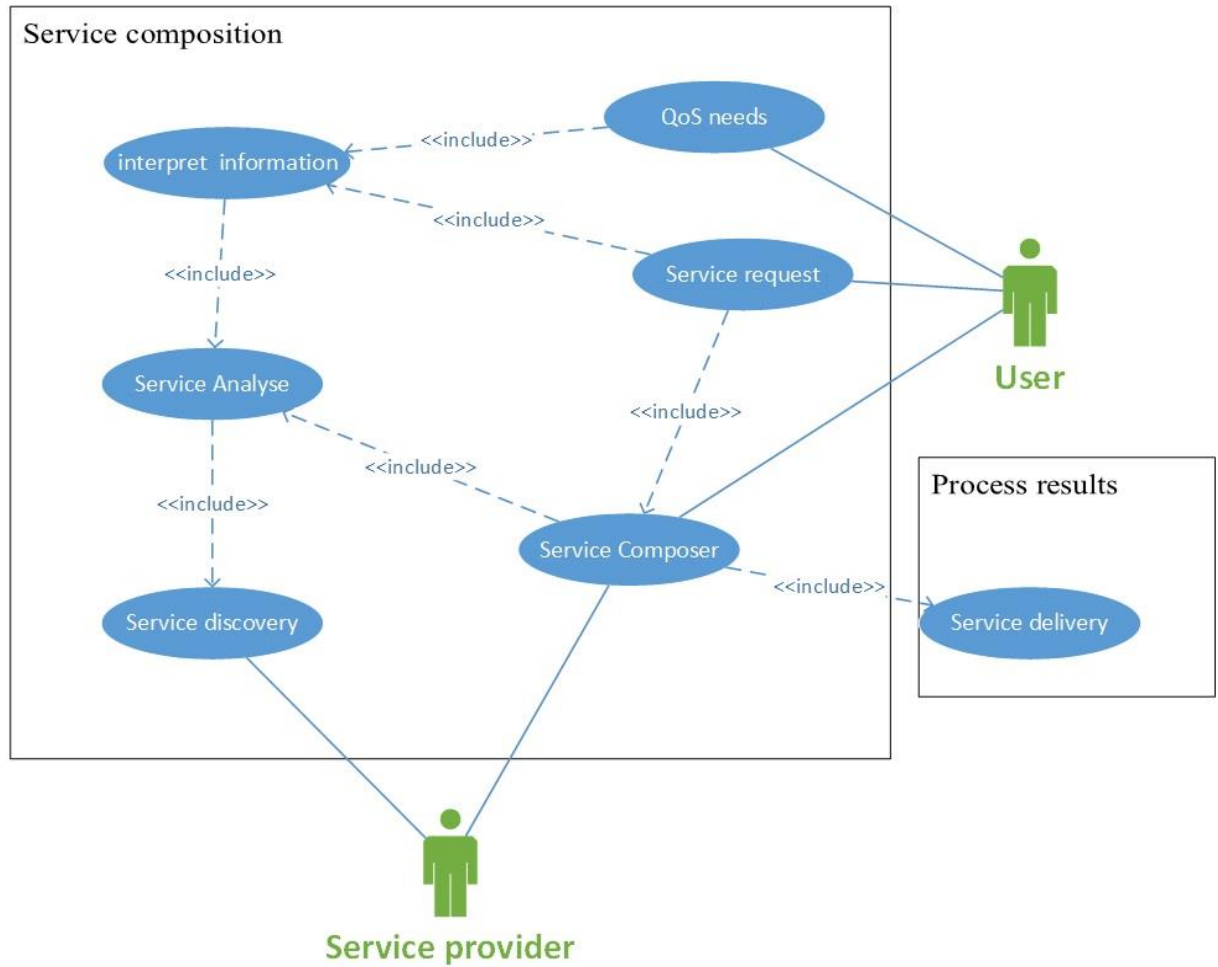

**Figure 2. Use case diagram of the service composition**

### 3.2.2. Evolutionary algorithms in the proposed method

The SFLA is a kind of memetic evolutionary algorithm that explores a globally optimal solution in a problem space [31]. The population in SFLA is constructed by frogs as memes, and each meme contains a number of memo types. The steps of the employed hybrid algorithm are displayed by the flowchart in Algorithm 1 in details. First, the initial parameters are set. Then, a specified number of memes form the initial population. After that, the ranks of them are calculated according to their fitness values. In the next stage, the memes will be sorted in the descending order based on their ranks. Afterward, the memes are distributed into a number of memeplexes in a random way one by one. Figure 3 illustrates the steps in the construction of the memeplexes. These memeplexes help to search the problem space in parallel in which the solutions in memeplexes are evolved by both local and global optimizations. Memes in each memeplex influence and evolves each other based on their ideas by using the GA operators i.e. two-point crossover and one point mutation. After the evolving process, memeplexes are shuffled to construct new populations for the next generation. These steps are repeated until a determined termination condition is achieved.

| **Algorithm 1**. The main function of the proposed hybrid SFGA: |
|---|
| Begin; |
| 1: Set the initial parameters |
| 2: Initialize the population with *n* frogs; |
| 3: For each frog, i∈n compute fitnesses (i); |
| 4: Rank the Frogs based on their fitness; |
| 5: Begin **while** the determined termination condition is achieved |
| 6:    Sort the frogs in the descending order according to their fitnesses; |
| 7:    Distribute the frogs into m memeplexes; |
| 8:    In each complex specify the best frogs via their fitnesses; |
| 9:       **Call** two-point crossover procedure to crossbred the best and worst frogs in the memeplexes; |
| 10:      **Call one-point** mutation procedure to mutate the randomly selected frogs; |
| 11:   Shuffle the evolved frogs; |
| 12: **End while** |
| 13: **Return** an optimized frog. |
| End; |

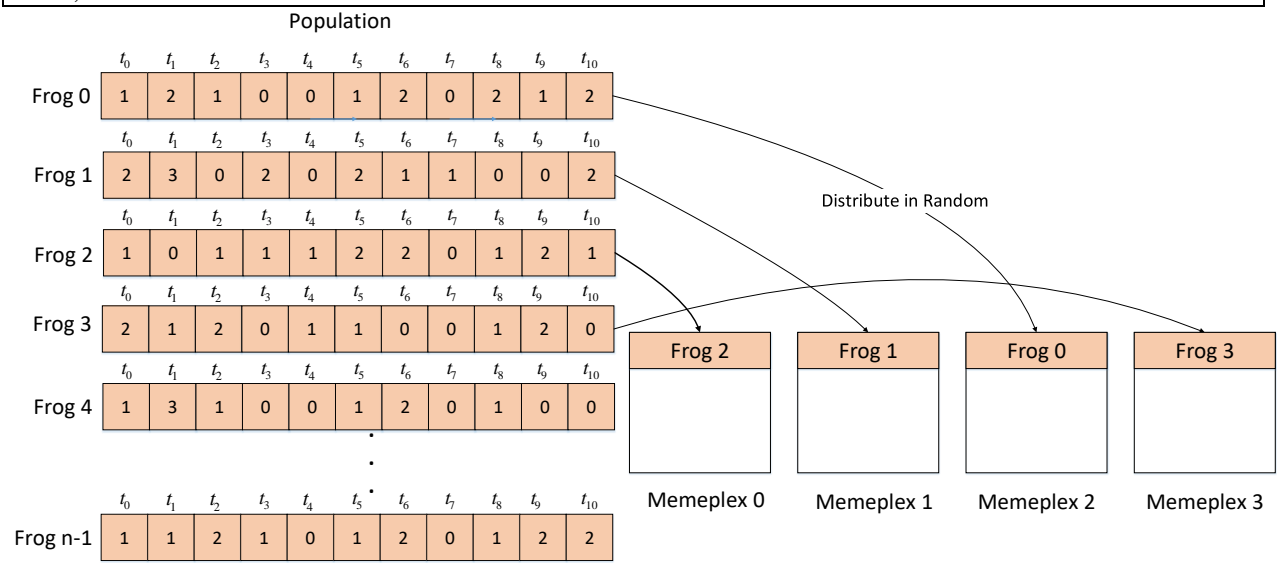

**Figure 3. Construction of the memeplexes**

A simple example of candidate services for the service composition problem is displayed in Table 1 and contains three aspects of QoS values: response time ($k_0$), Energy ($k_1$), and cost ($k_2$). For instance, values 30, 78 and 90 in the candidate service $W_{0,0}$ for $t_0$ are related to $k_0$, $k_1$, and $k_2$ respectively.

| Table 1. Example of quality matrix and properties | | | | |
|---|---|---|---|---|
| **Quality matrix** | | **$k_0$** | **$k_1$** | **$k_2$** |
| $t_0$ | $w_{0,0}$ | 30 | 48 | 90 |
| | $w_{0,1}$ | 26 | 70 | 40 |
| | $w_{0,2}$ | 19 | 96 | 63 |
| $t_1$ | $w_{1,0}$ | 65 | 100 | 49 |
| | $w_{1,1}$ | 38 | 79 | 70 |
| | $w_{1,2}$ | 55 | 89 | 60 |
| | $w_{1,3}$ | 67 | 99 | 41 |
| $t_2$ | $w_{2,0}$ | 46 | 114 | 96 |
| | $w_{2,1}$ | 68 | 125 | 76 |
| | $w_{2,2}$ | 90 | 111 | 47 |
| $t_3$ | $w_{3,0}$ | 69 | 116 | 57 |
| | $w_{3,1}$ | 87 | 99 | 86 |
| | $w_{3,2}$ | 46 | 147 | 39 |
| $t_4$ | $w_{4,0}$ | 74 | 117 | 91 |
| | $w_{4,1}$ | 61 | 86 | 45 |
| $t_5$ | $w_{5,0}$ | 29 | 109 | 88 |
| | $w_{5,1}$ | 40 | 90 | 37 |
| | $w_{5,2}$ | 63 | 120 | 101 |
| $t_6$ | $w_{6,0}$ | 74 | 71 | 44 |
| | $w_{6,1}$ | 39 | 113 | 93 |
| | $w_{6,2}$ | 45 | 110 | 73 |
| $t_7$ | $w_{7,0}$ | 61 | 100 | 28 |
| | $w_{7,1}$ | 49 | 98 | 74 |
| $t_8$ | $w_{8,0}$ | 66 | 130 | 55 |
| | $w_{8,1}$ | 52 | 82 | 36 |
| | $w_{8,2}$ | 73 | 121 | 105 |
| $t_9$ | $w_{9,0}$ | 80 | 33 | 58 |
| | $w_{9,1}$ | 37 | 105 | 51 |
| $t_{10}$ | $w_{10,0}$ | 29 | 79 | 87 |
| | $w_{10,1}$ | 74 | 75 | 42 |
| | $W_{10,2}$ | 54 | 77 | 106 |

For all the tasks in Table 1 from 0 to 10, the proper services among the candidates should be selected. Hence, to optimize the solutions, the two-point crossover and mutation operators are embedded in the SFLA in the proposed algorithm. In the crossover function, the best and worst solutions in memeplexes are crossbred to have local optimization otherwise the operator will be executed between the worst solution and global best to have global optimization. Figure 4 shows the sample of the crossover function in the suggested algorithm. Also, Algorithm 2 describes the applied Two-point Crossover procedure.

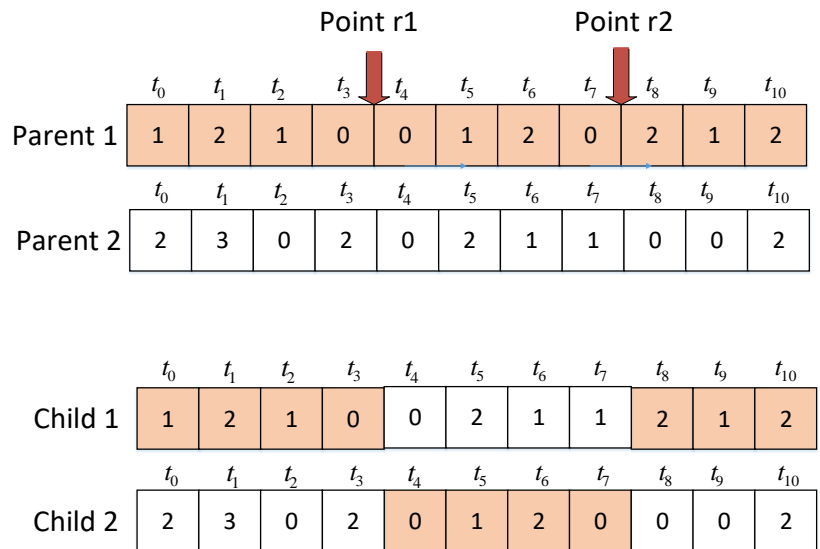

Figure 4. Two-point crossover

**Algorithm 2**. Two-point Crossover procedure

1: Begin;
2: **for each** complex **do**
3:   select the fittest frog from complex i as parent 1;
4:   select the worst frog from the complex i ;
5:   generate two random crossover points r1 and r2;
6:   swap the interval segments of the two selected points generate two new children i.e. child 1 and child 2 ;
7:   compute the fitnesses of the child 1 and child 2 ;
8:   **if** the fitness of worst frog<= max(child 1, child 2 ) **then**
9:     **replace** the worst frog with max (child 1, child 2)**:**
10:    end if
11:  **else**
12:    crossbred the worst frog with the global best frog to generate two new children i.e. child 1 and child 2 ;
13:    compute the fitnesses of the child 1 and child 2;
14:    **if** the fitness of worst frog<= max(child 1, child 2 ) **then**
15:      **replace** the worst frog with max (child 1, child 2)**:**
16:      end if
17:      else
18:        **replace** the worst frog with a new random generated frog;
19: end for
20: **end for each**

```
  End;
```

In the next phase of the GA in the proposed algorithm, a random solution is selected to mutate with a single point mutation operator. As shown in Figure 5, a random memotype is chosen and then replaced with a new random one which is generated among the related candidate services. This process prevents the solutions from trapping into local minima. Moreover, the exploited one-point mutation operator is detailed in Algorithm 3.

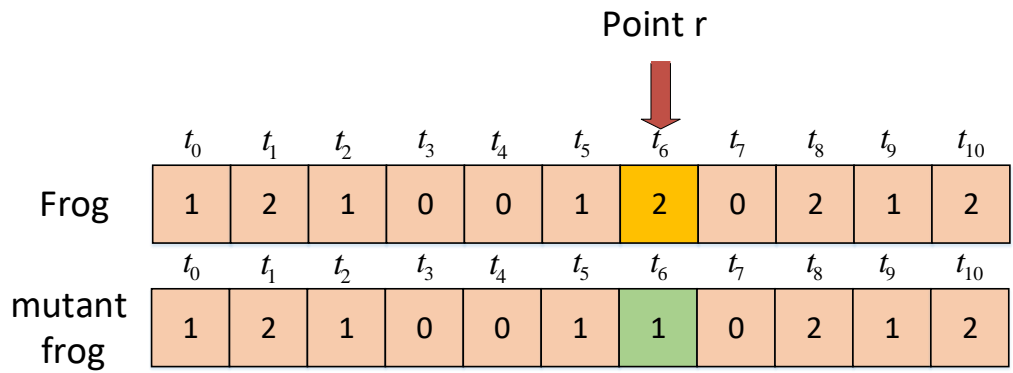

Figure 5. The single point mutation process

**Algorithm 3**. One-point mutation procedure

```
   Begin;
1: for each complex do
2:   for i=0 to (size of complex ×30/100) do // probability is 0.3
3:     select a random frog in complex i;// except the best frog
4:     generate a random point r;
5:     generate a random number among the candidate services;
6:     exchange the value of the selected memotype with the generated candidate service number;
7:     compute the fitness of the obtained new frog;
8:      replace the mutant frog;
9:   end for
10: end for each
   End;
```

### 3.3. Fitness function

Three QoS factors including response time, energy, and cost are important factors for selecting the optimal service composition due to the battery life of mobile devices in a mobile cloud computing environment. The QoS parameters are specified in positive and negative factors. Increasing values for positive factors, such as availability and throughput are useful for users. However, decreasing values for negative factors, such as energy and response time is helpful for users. Increasing positive criteria and decreasing negative criteria are promoted by the fitness function [32].

According to the existing QoS parameters as the negative factors, the *Response time, Energy* and *Cost* have a negative influence. So, the fitness function is shown as follows:

$$Fitness = W_1 * Responsetime + W_2 * Energy + W_3 * Cost \qquad (1)$$

Where $w_1$, $w_2$, and $w_3$ are positive weights and $\sum_{i=1}^{4} w_i = 1$. To evaluate the fitness value, the main objective of the proposed algorithm is selecting a composited service with a minimized fitness value. According to Table 2, the existing QoS factors are calculated with sum and product operations [33] and are normalized between (0, 1).

Table 2. QoS aggregation formulas for sequential composition model [34].

| QoS property | Sequence | Circle | Branch | Fork |
|---|---|---|---|---|
| Response Time (T) | $\sum_{i=1}^{m} T(a_i)$ | $k \cdot \sum_{i=1}^{n} T(a_i)$ | $\sum_{i=1}^{m} P_i \cdot T(s_i^b)$ | $max_{i=1}^{p} T(s_i^f)$ |
| Energy (E) | $\sum_{i=1}^{m} E(a_i)$ | $k \cdot \sum_{i=1}^{n} E(a_i)$ | $\sum_{i=1}^{m} P_i \cdot E(s_i^b)$ | $max_{i=1}^{p} E(s_i^f)$ |
| Cost (C) | $\prod_{i=}^{m} C(a_i)$ | $(\prod_{i=1}^{n} C(a_i))^k$ | $\sum_{i=1}^{m} P_i \cdot C(s_i^b)$ | $min_{i=1}^{p} C(s_i^f)$ |

## 4. Experimental results

In this section, the QWS that was created by Al-Masri [35] that contains 2507 real web services and we considered 3 features of QoS factors for this experiment. To analyze the results of the proposed algorithm, Culture Algorithm (CA), Genetic Algorithm (GA), Particle Swarm Optimization (PSO) and GAPSO algorithm are used in 3 different scenarios. To have a fair comparison, the mentioned algorithms are selected in the experiments because of their natures as evolutionary algorithms. Also, the applied methods are programmed and simulated by C# language in Visual Studio 2017 environment.

According to Table 3, the proposed scenarios are considered. The response time, energy and cost factors of the proposed algorithm with respect to other algorithms are compared using different values for the number of services.

Table 3. Details of scenarios used in experiments

| Scenarios | Number of services | Goal |
|---|---|---|
| First Scenario | 10, 20, 30, 40, 50 | Evaluating response time, energy and cost in existing algorithms |
| Second scenario | 100, 200, 300, 400 | |

## 4.1 Analyzing the first scenario

In this scenario, the proposed algorithm is compared with four other algorithms. Also, the number of services lets 10, 20, 30, 40 and 50 for each iteration.

In Figure 6, the cost factor of the PSO algorithm is more than four other algorithms. In a comparison between existing algorithms, the SFGA method reduces the service composition cost reasonably. When the number of requests increases, the cost decreases significantly. The reason for reducing the cost of service and its composition is the optimal selection of services and better decisions of the proposed algorithm.

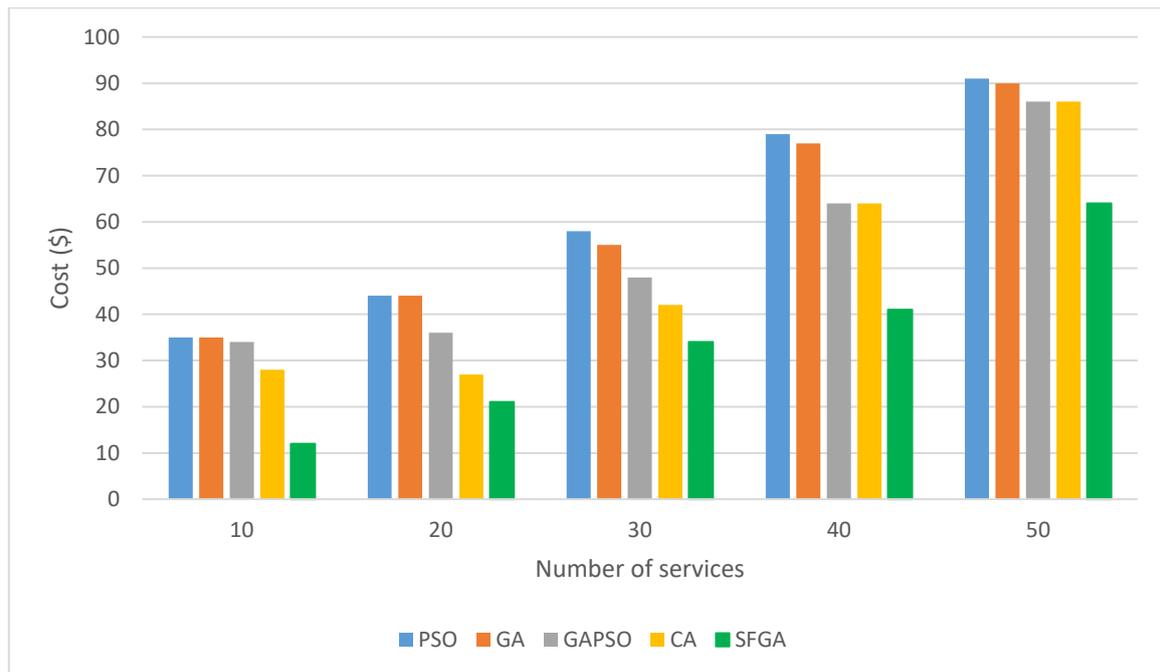

**Figure 6. Evaluation of the cost factor for existing algorithms in the first scenario.**

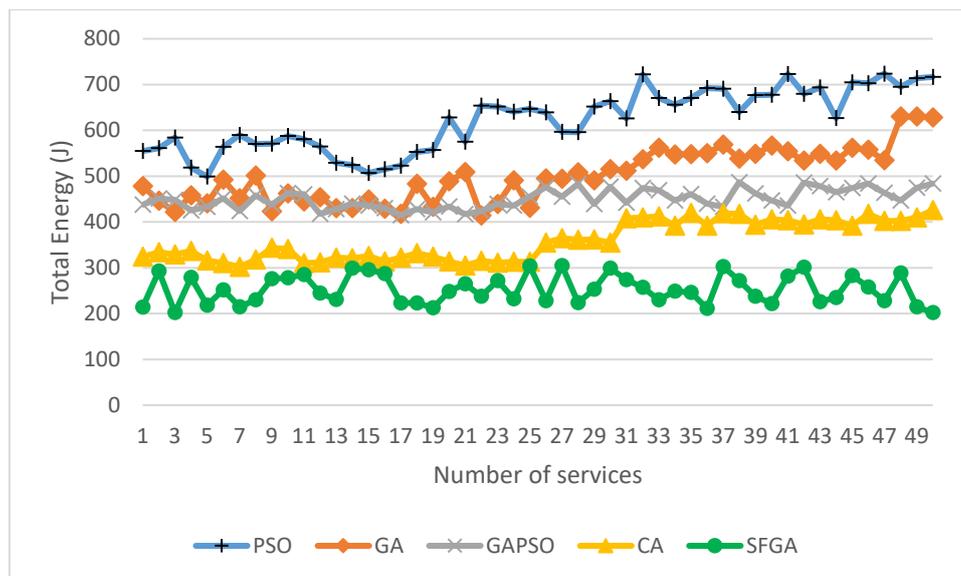

**Figure 7. Evaluation of the total energy consumption for the service composition approach in the first scenario**.

Figure 7 shows the total energy of mobile services in the process of the final service composition. According to Figure 7, the total energy of a set of mobile service providers in the algorithm SFGA is lower than the other algorithms, which indicates that this algorithm provides the more appropriate distribution of requests with minimum energy consumption for mobile service providers.

In addition, the total energy consumption values by the methods from figure 7 are plotted by the boxplot tool is the SPSS environment which is shown in Figure 8. According to the boxplot, it is obvious that the SFGA approach betters the other algorithms in the case of energy consumption. Also, the mean values of the services for energy consumption are depicted in Table 4. The results illustrate that the proposed method is undoubtedly minimized the mentioned algorithms.

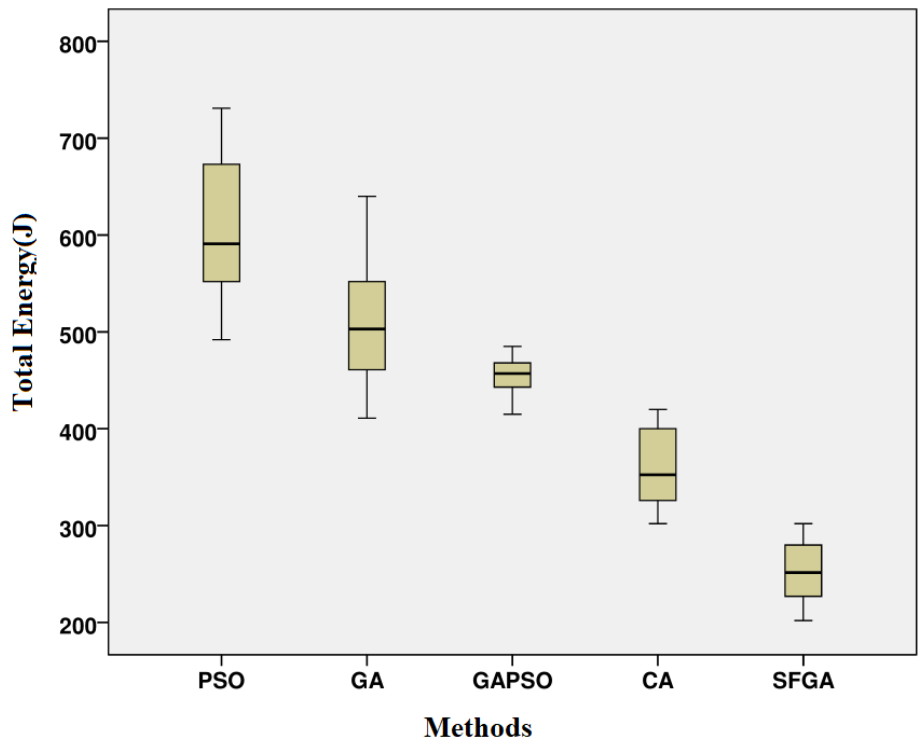

**Figure 8. Boxplot of the total energy consumption values by the methods from Figure 7**

**Table 4.** The mean values of Total Energy results for the methods in Figure 7

|  | PSO | GA | GAPSO | CA | SFGA |
|---|---|---|---|---|---|
| **Mean** | 609.0000 | 506.3400 | 454.5000 | 359.7000 | 251.7400 |
| **Std. Deviation** | 69.49967 | 55.46008 | 19.77346 | 40.16052 | 30.10716 |

Finally, Figure 9 depicts the response time factor for existing meta-heuristic algorithms based on the first scenario. According to this figure, the SFGA algorithm has outperformed results in this scenario when the number of services is increased with minimum results.

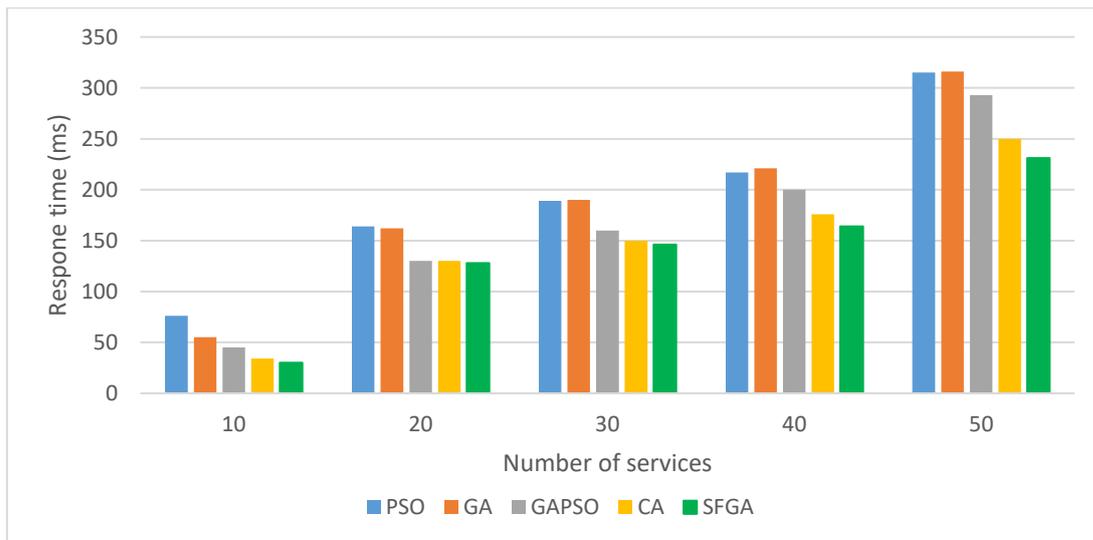

**Figure 9. Evaluation of the response time for the service composition approach in the first scenario.**

## 4.2 Evaluation of the second scenario

In the second scenario, we assume that the number of services is considered 100, 200, 300, and 400 on the QWS. In Figure 10, the cost factor of the SFGA method is lower than the other algorithms. When the number of requests increases, the cost decreases significantly. The reason for reducing the cost of service and its composition is the optimal selection of services and better decisions of the proposed algorithm.

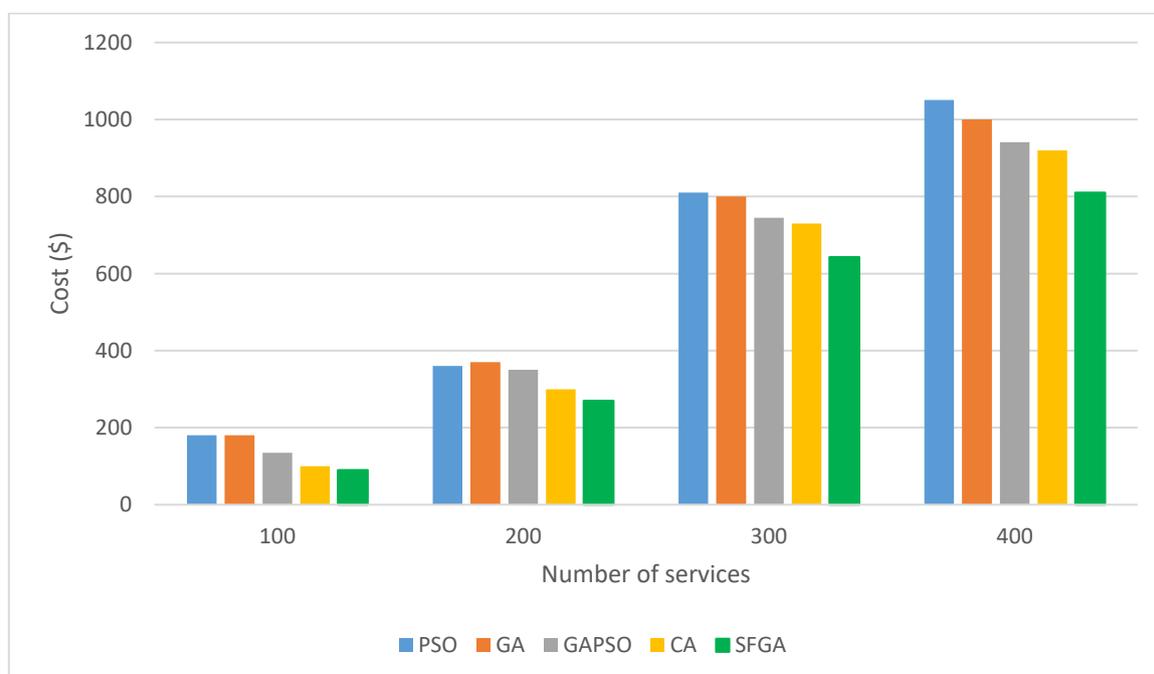

**Figure 10. Evaluation of the cost factor for existing algorithms in the second scenario.**

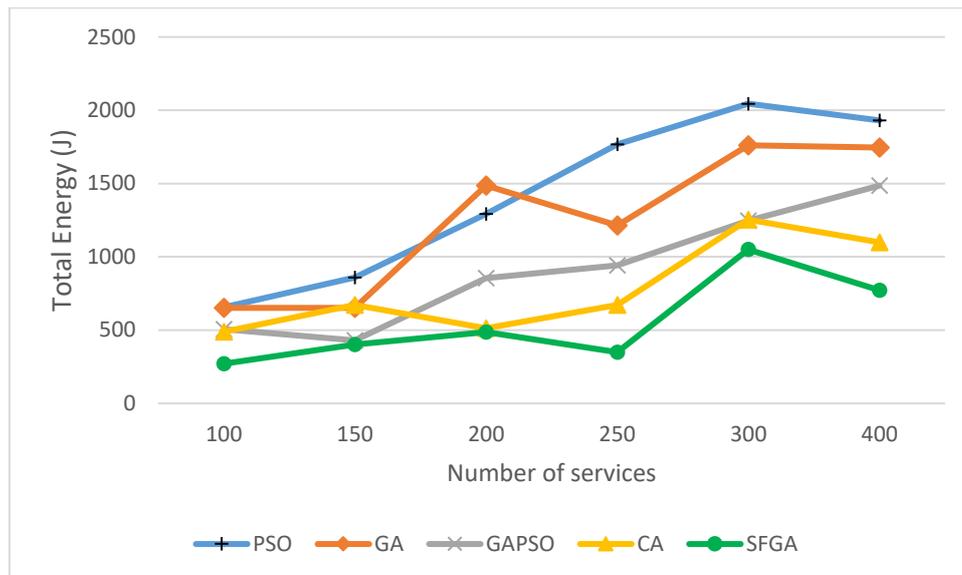

**Figure 11. Evaluation of the total energy consumption for the service composition approach in the second scenario.**

Figure 11 shows the total energy consumption of mobile services in the process of the final service composition. According to Figure 11, the total energy of a set of mobile service providers in the algorithm SFGA is lower than the other algorithms, which indicates that this algorithm provides the more appropriate distribution of requests with minimum energy consumption for mobile service providers.

Furthermore, the results of energy consumption by the algorithms from Figure 11 are represented by the boxplot in Figure 12. Also, Table 5 shows the PSO, GA, GAPSO, CA, and SFGA average values of energy consumption from services in Figure 11. The plotted boxes and the mean values reveal that the proposed algorithm significantly outperforms the other methods in terms of energy saving.

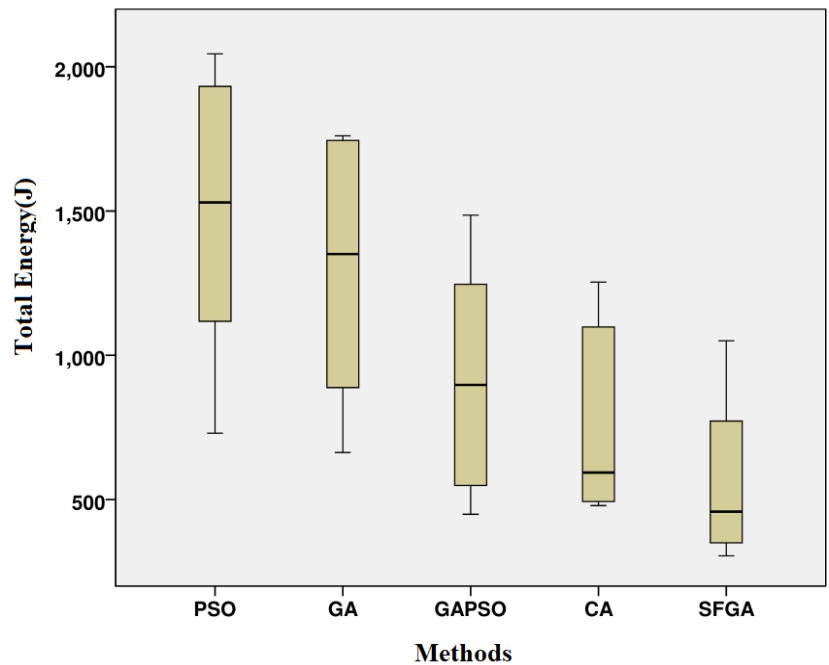

**Figure 12. Boxplot of the total energy consumption values by the methods from Figure 11**

**Table 5.** The mean values of Total Energy results for the methods in Figure 11

|  | **PSO** | **GA** | **GAPSO** | **CA** | **SFGA** |
|---|---|---|---|---|---|
| **Mean** | 1424.8333 | 1267.3333 | 951.1667 | 770.3333 | 570.8333 |
| **Std. Deviation** | 580.09772 | 484.63292 | 361.28903 | 326.47001 | 290.98620 |

Finally, Figure 13 depicts the response time factor for existing meta-heuristic algorithms based on the first scenario. According to this figure, the SFGA algorithm has outperformed results in this scenario when the number of services is increased with minimum results.

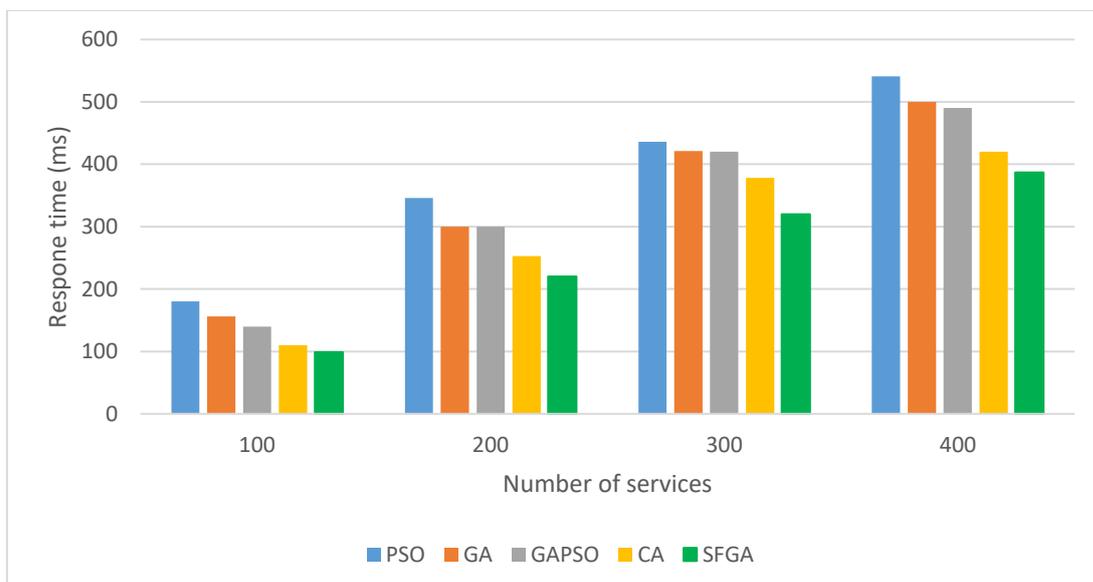

Figure 13. Evaluation of the response time for the service composition approach in the second scenario.

# 5. Conclusion

Service combination is one of the five steps introduced in the service-oriented architecture model. Other steps include service identification, service description, service objectification, service composition, and service implementation. In the present research, we only focused on service composition and its methods. Many efforts have been undertaken to combine web services in the mobile cloud computing environment that any of the methods outlined has its advantages and disadvantages. In the proposed method, quality measurement factors have been divided into three negative factors (response time, energy, and cost). Using the SFGA algorithm resulted in faster service selection and thus better service composition that can be specifically effective in response time and service cost.

In this paper, the proposed hybrid service composition method in mobile cloud computing was compared by four methods, namely PSO, GA, and PSOGA and CA algorithms then it was investigated by three QoS criteria. The results show that the SFGA algorithm, compared to others, decreases the service composition cost, energy consumption, and the response time to requests, as three important factors in improving the quality of service. Furthermore, more QoS criteria for the service composition problem will be considered in our future work. Also, some new meta-heuristic algorithms, such as WOA-BAT Algorithm [36], Donkey and Smuggler Optimisation Algorithm [37], Fitness Dependent Optimiser [38], and Modified Grey Wolf Optimiser [39] can be applied to evaluate the energy-aware service composition approach in future work.

## References

[1] N. Tariq, M. Asim, Z. Maamar, M. Z. Farooqi, N. Faci, and T. Baker, "A Mobile Code-driven Trust Mechanism for detecting internal attacks in sensor node-powered IoT," Journal of Parallel and Distributed Computing, vol. 134, pp. 198-206, 2019/12/01/ 2019.

[2] F. Safara, A. Souri, T. Baker, I. Al Ridhawi, and M. Aloqaily, "PriNergy: a priority-based energy-efficient routing method for IoT systems," The Journal of Supercomputing, 2020/01/16 2020.

[3] M. Dighriri, A. S. D. Alfoudi, G. M. Lee, T. Baker, and R. Pereira, "Comparison Data Traffic Scheduling Techniques for Classifying QoS over 5G Mobile Networks," in 2017 31st International Conference on Advanced Information Networking and Applications Workshops (WAINA), 2017, pp. 492-497.

[4] Y. Ma, S. Wang, P. C. K. Hung, C. H. Hsu, Q. Sun, and F. Yang, "A Highly Accurate Prediction Algorithm for Unknown Web Service QoS Values," IEEE Transactions on Services Computing, vol. 9, pp. 511-523, 2016.

[5] M. Dighriri, G. M. Lee, and T. Baker, "Measurement and Classification of Smart Systems Data Traffic Over 5G Mobile Networks," in Technology for Smart Futures, M. Dastbaz, H. Arabnia, and B. Akhgar, Eds., ed Cham: Springer International Publishing, 2018, pp. 195-217.

[6] M. Asim, A. Yautsiukhin, A. D. Brucker, T. Baker, Q. Shi, and B. Lempereur, "Security policy monitoring of BPMN-based service compositions," Journal of Software: Evolution and Process, vol. 30, p. e1944, 2018/09/01 2018.

[7] T. Baker, D. Lamb, A. Taleb-Bendiab, and D. Al-Jumeily, "Facilitating Semantic Adaptation of Web Services at Runtime Using a Meta-Data Layer," in 2010 Developments in E-systems Engineering, 2010, pp. 231-236.

[8] H. Ma, H. Zhu, Z. Hu, W. Tang, and P. Dong, "Multi-valued collaborative QoS prediction for cloud service via time series analysis," Future Generation Computer Systems, vol. 68, pp. 275-288, 2017/03/01/ 2017.

[9] Z. Ali, L. Jiao, T. Baker, G. Abbas, Z. H. Abbas, and S. Khaf, "A Deep Learning Approach for Energy Efficient Computational Offloading in Mobile Edge Computing," IEEE Access, vol. 7, pp. 149623-149633, 2019.

[10] A. Souri, A. Hussien, M. Hoseyninezhad, and M. Norouzi, "A systematic review of IoT communication strategies for an efficient smart environment," Transactions on Emerging Telecommunications Technologies, vol. n/a, p. e3736, 2019/08/29 2019.

[11] M. Eusuff, K. Lansey, and F. Pasha, "Shuffled frog-leaping algorithm: a memetic meta-heuristic for discrete optimization," Engineering Optimization, vol. 38, pp. 129-154, 2006/03/01 2006.

[12] J. H. Holland, Adaptation in natural and artificial systems : an introductory analysis with applications to biology, control, and artificial intelligence. Ann Arbor: University of Michigan Press, 1975.

[13] J. Parejo, P. Fernandez, and A. Ruiz-Cortés, QoS-aware services composition using tabu search and hybrid genetic algorithms vol. 2, 2008.

[14] S.-M. Huang, Y.-T. Chu, S.-H. Li, and D. C. Yen, "Enhancing conflict detecting mechanism for Web Services composition: A business process flow model transformation approach," Information and Software Technology, vol. 50, pp. 1069-1087, 2008/10/01/ 2008.

[15] F. Mardukhi, N. NematBakhsh, K. Zamanifar, and A. Barati, "QoS decomposition for service composition using genetic algorithm," Applied Soft Computing, vol. 13, pp. 3409-3421, 2013/07/01/ 2013.

[16] G. H. Alférez, V. Pelechano, R. Mazo, C. Salinesi, and D. Diaz, "Dynamic adaptation of service compositions with variability models," Journal of Systems and Software, vol. 91, pp. 24-47, 2014/05/01/ 2014.

[17] K. Su, L. Ma, X. Guo, and Y. Sun, An Efficient Discrete Invasive Weed Optimization Algorithm for Web Services Selection vol. 9, 2014.

[18]   R. Karim, C. Ding, A. Miri, and M. S. Rahman, "Incorporating service and user information and latent features to predict QoS for selecting and recommending cloud service compositions," Cluster Computing, vol. 19, pp. 1227-1242, September 01 2016.

[19]   A. Ramírez, J. A. Parejo, J. R. Romero, S. Segura, and A. Ruiz-Cortés, "Evolutionary composition of QoS-aware web services: A many-objective perspective," Expert Systems with Applications, vol. 72, pp. 357-370, 2017/04/15/ 2017.

[20]   A. Souri, A. M. Rahmani, and N. Jafari Navimipour, "Formal verification approaches in the web service composition: a comprehensive analysis of the current challenges for future research," International Journal of Communication Systems, vol. 31, p. e3808, 2018.

[21]   A. Souri, A. M. Rahmani, N. J. Navimipour, and R. Rezaei, "A hybrid formal verification approach for QoS-aware multi-cloud service composition," Cluster Computing, pp. 1-18, 2019.

[22]   M. Sun, Z. Zhou, J. Wang, C. Du, and W. Gaaloul, "Energy-Efficient IoT Service Composition for Concurrent Timed Applications," Future Generation Computer Systems, vol. 100, pp. 1017-1030, 2019.

[23]   D. Ardagna and B. Pernici, "Adaptive service composition in flexible processes," IEEE Transactions on software engineering, vol. 33, 2007.

[24]   Y. Huo, Y. Zhuang, J. Gu, S. Ni, and Y. Xue, "Discrete gbest-guided artificial bee colony algorithm for cloud service composition," Applied Intelligence, vol. 42, pp. 661-678, 2015.

[25]   C. Jian, M. Li, and X. Kuang, "Edge cloud computing service composition based on modified bird swarm optimization in the internet of things," Cluster Computing, 2018/01/08 2018.

[26]   S. Asghari and N. J. Navimipour, "Nature inspired meta-heuristic algorithms for solving the service composition problem in the cloud environments," International Journal of Communication Systems, p. e3708, 2018.

[27]   A. Naseri and N. J. Navimipour, "A new agent-based method for QoS-aware cloud service composition using particle swarm optimization algorithm," Journal of Ambient Intelligence and Humanized Computing, pp. 1-14, 2018.

[28]   C. Ben Njima, Y. Gamha, C. Ghedira Guegan, and L. Ben Romdhane, "Development of a mobile web services discovery and composition model," Cluster Computing, 2019/01/18 2019.

[29]   J. Luo, X. Li, and M.-r. Chen, "A novel hybrid shuffled frog leaping algorithm for vehicle routing problem with time windows," Information Sciences, vol. 316, 04/01 2015.

[30]   A. Jula, E. Sundararajan, and Z. Othman, "Cloud computing service composition: A systematic literature review," Expert systems with applications, vol. 41, pp. 3809-3824, 2014.

[31]   M. M. Eusuff and K. E. Lansey, "Optimization of water distribution network design using the frog leaping algorithm," J. Water. Res. Plan. Manage., vol. 129, pp. 10-25, 01/01 2003.

[32]   A. Souri, M. Rahmani Amir, J. Navimipour Nima, and R. Rezaei, "Formal modeling and verification of a service composition approach in the social customer relationship management system," Information Technology & People, vol. 32, pp. 1591-1607, 2019.

**Godar J. Ibrahim, Tarik A. Rashid, Mobayode O. Akinsolu. An energy efficient service composition mechanism using a hybrid meta-heuristic algorithm in a mobile cloud environment, Journal of Parallel and Distributed Computing, DOI: 10.1016/j.jpdc.2020.05.002**
[33]    M. R. Mesbahi, A. M. Rahmani, and M. Hosseinzadeh, "Reliability and high availability in cloud computing environments: a reference roadmap," Human-centric Computing and Information Sciences, vol. 8, p. 20, 2018.

[34]    M. E. Khanouche, H. Gadouche, Z. Farah, and A. Tari, "Flexible QoS-aware services composition for service computing environments," Computer Networks, vol. 166, p. 106982, 2020.

[35]    E. Al-Masri and Q. H. Mahmoud, "Investigating web services on the world wide web," in Proceedings of the 17th international conference on World Wide Web, 2008, pp. 795-804.

[36] H. Mohammed, S. Umar, and T. Rashid. A Systematic and Meta-Analysis Survey of Whale Optimization Algorithm, Computational Intelligence and Neuroscience, (2019), Article ID 8718571, 25 pages.https://doi.org/10.1155/2019/8718571

[37] A. Shamsaldin, T. Rashid, R. Al-Rashid Agha, N. Al-Salihi, M. Mohammadi . Donkey and Smuggler Optimization Algorithm: A Collaborative Working Approach to Path Finding, Journal of Computational Design and Engineering. 6 (4), (2019), Pages 562-583.
DOI: https://doi.org/10.1016/j.jcde.2019.04.004

[38] J. Abdullah and T. Ahmed. Fitness Dependent Optimizer: Inspired by the Bee Swarming Reproductive Process," in IEEE Access 7 (2019) pp. 43473-43486. DOI:https://doi.org/10.1109/ACCESS.2019.2907012

[39] T. Rashid, D. Abbas, Y. Turel . A multi hidden recurrent neural network with a modified grey wolf optimizer. PLoS ONE 14(3), (2019), e0213237.
DOI:  https://doi.org/10.1371/journal.pone.0213237

24